\documentclass{nature}
\usepackage{amsthm,amsmath,amssymb}
\usepackage{lineno}
\usepackage{xcolor}
\usepackage[figuresleft]{rotating}
\usepackage{hyperref}
\usepackage{graphicx}
\usepackage{longtable}
\usepackage{float} 

\def\@citen#1#2{\mbox{\scriptsize #1\if@tempswa , #2\fi}}

\makeatletter
\let\saved@includegraphics\includegraphics
\AtBeginDocument{\let\includegraphics\saved@includegraphics}
\renewenvironment*{figure}{\@float{figure}}{\end@float}
\makeatother

\newcommand{\apj}{Astrophys. J.}

\newcommand{\pasp}{Publ. Astron. Soc. Pac.}

\newcommand{\araa}{Annu. Rev. Astron. Astrophys.}
\newcommand{\mnras}{Mon. Not. R. Astron. Soc.}
\newcommand{\apjl}{Astrophys. J. Let.}
\newcommand{\aap}{Astron. Astrophys.}
\newcommand{\aj}{Astron. J.}

\newcommand{\prl}{Phys. Rev. Lett.}
\newcommand{\prd}{Phys. Rev. D}

\newcommand{\ssr}{Space. Sci. Reviews.}

\newcommand{\jcap}{Journal of Cosmology and Astroparticle Physics}

\title{High-energy neutrino emission from the Type~IIn supernova SN~2017hcd}

\author{Shunhao Ji$^{1,\ast}$, Zhongxiang Wang$^{1,\ast}$, Litao Zhu$^{1}$ \& Dong Zheng$^{1}$}

\begin{document}

\maketitle

\begin{affiliations}
\item Department of Astronomy, School of Physics and Astronomy, Key Laboratory of Astroparticle Physics of Yunnan Province, Yunnan University, Kunming 650091, China
\end{affiliations}


\begin{abstract}
Neutrino astronomy provides another window to exploring the Universe, 
exemplified by the detection of a megaelectronvolt neutrino burst
from the core-collapse supernova (CCSN) SN~1987A 
(refs.~\citenum{hir+87,bio+87}).  Commonly discussed 
theories suggest that some CCSNe could produce 
neutrinos with energies a thousand times more than those of 
SN~1987A \cite{tm18}, which has been probed with new-generation 
facilities \cite{abb+12,aar+15,abb+23}.  
The interaction of SN ejecta with a dense circumstellar medium (CSM) or a jet, 
launched in a CCSN, being choked in the stellar envelope of the progenitor or 
an outside CSM are both well-accepted scenarios for the high-energy neutrino
production. Here we report the detection of a high-energy neutrino flare at a 
3.9$\sigma$ significance from SN~2017hcd, made by our analysis of the public 
track-like neutrino data taken by the IceCube Neutrino 
Observatory \cite{IceCube17}. 
A Type IIn SN with optical emissions arising from the ejecta--CSM interaction,
	SN~2017hcd's neutrino flare lasted $\sim$1--2 month,
with its central time $\sim$14-day prior to the SN's optical discovery time.
Its estimated isotropic neutrino energy (all flavors) 
is approximately two orders of magnitude higher than 
	the energy ($\sim 10^{50}$\,erg)
carried in the SN's ejecta, too high to be explained with 
the ejecta--CSM scenario.
Thus, a choked jet may be the source of the neutrino flare.
\end{abstract}



It has been nearly 40 years since the nearby supernova (SN) explosion 
SN 1987A, which occurred in the Large Magellanic Cloud and whose burst of 
megaelectronvolt (MeV) neutrinos 
were successfully detected \cite{hir+87,bio+87}. While new-generation
neutrino detectors are eagerly waiting to detect such events from similar
core-collapse SNe (CCSNe) that are expected to occur once every $\sim$30 years
in the Milky Way \cite{hk18}, the
IceCube neutrino observatory \cite{IceCube17} at the South Pole, as well
as several other detectors, have opened a
multi-messenger window for detecting gigaelectronvolt (GeV) to 
teraelectronvolt (TeV) neutrinos emitted from astrophysical objects
\cite{txs0506b,txs0506a,ngc1068,galneu},
and have since been used to search for high-energy neutrinos emitted from
CCSNe \cite{ice+11,abb+12,aar+15,smm18,abb+23,cha+24,ste+25}.

The CCSNe, which mark the deaths of massive stars, are among the most luminous 
transient phenomena in the Universe. In addition to the burst of MeV neutrinos 
produced in their core-collapse process \cite{hk18}, it has been
widely discussed that when there is a substantially dense circumstellar 
medium (CSM) around a CCSN, which is present due to the progenitor's 
significant mass loss in the years before the explosion \cite{ofe+14,ran+21},
high-energy Gev--TeV neutrinos would also be produced through 
the interaction of the SN's ejecta with the CSM \cite{tm18}. As a subclass of 
Type II SNe, the Type IIn SNe (SNe~IIn) are characterized by optical 
spectra with narrow Balmer emission lines, an indication of shock interaction 
between the SN ejecta and the CSM.  SNe~IIn are thus considered as particularly 
promising targets for detecting high-energy neutrinos from CCSNe. 

\begin{figure}[t]
    \centering
	\includegraphics[width=0.97\textwidth]{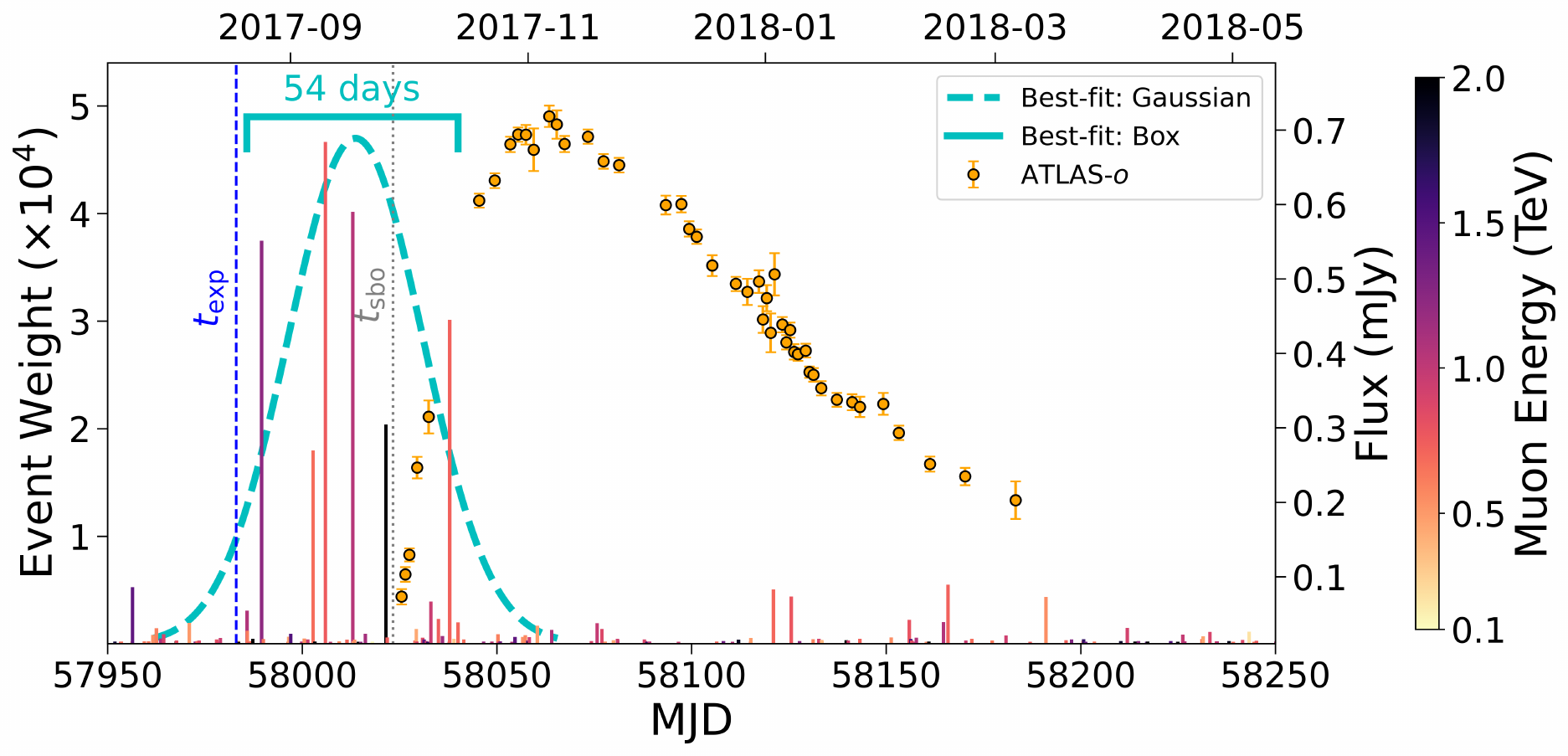}
	\caption{\textbf{Neutrino flare determined at SN~2017hcd's position.}
The colored vertical lines are the individual neutrino events detected, with
	their heights being the weights (calculated with $\gamma = 3.5$; see 
	section `IceCube data analysis') and their colors being
	the reconstructed muon energies (indicated by the right-side color bar).
	The time durations of the flare, from the box and Gaussian
	time profile (marked by the 54-days cyan line and cyan dashed curve 
	respectively)
	are shown.  The whole X-axis time duration is the \textit{a priori} 
	window $T_{\rm win}$, within which the flare was searched.
	The central times of the durations precede the shock breakout
	time (marked by the gray dotted line) of the SN by $\sim$10\,days; 
	the latter was determined from the ATLAS o-band light curve 
	(orange dots; see section `Photometric data and light-curve analysis').
	The estimated explosion time of the SN is marked by a blue dashed line.}
	\label{fig:lc}
\end{figure}

SN~2017hcd was discovered by the Asteroid Terrestrial-impact Last Alert System 
(ATLAS) \cite{Tonry+18} on October~1 2017 (MJD 58027.48) at the position of
R.A. = 25$^\circ$.71604, Decl. = +31$^\circ$.48238 (J2000).
It was classified as an SN IIn with a redshift $z$ of 0.035
(see section `Photometric data and light-curve analysis'), 
which corresponds to a luminosity distance of 159.2 megaparsec (Mpc). It is 
located at the edge of the nearby galaxy MCG+05-05-002 
(see Extended Data Fig.~\ref{fig:image}) and should be
associated with this galaxy given that they both have the same redshift value.
The optical light curves of SN~2017hcd (Fig.~\ref{fig:lc} and Extended Data
Fig.~\ref{fig:fit}), although only 
mainly at ATLAS $o$ band, are generally
similar to those of SNe IIn \cite{mor+14,rv25}, and we
estimated a total radiation energy of $\sim 10^{50}$\,erg from analyzing
the light curves (see section~`Photometric data and light-curve analysis').
An optical spectrum of SN~2017hcd was taken 16 days after 
the discovery, and it also shows features typical of SNe IIn
(Extended Data Fig.~\ref{fig:spec}; see section `Optical spectrum analysis'). 

We searched for neutrino emissions from SN~2017hcd 
by performing a time-dependent unbinned maximum likelihood 
analysis 
to the public data provided by IceCube, 
which are implemented in the {\tt SkyLLH} framework released by 
IceCube (see section `IceCube data analysis'). 
An \textit{a~priori} time window T$_{\rm win}$ for the search was set to 
be 300\,days,
from MJD~57950 to MJD~58250 (Fig.~\ref{fig:lc}). This window was such
that not only was the property of SNe IIn considered, whose
shock breakout (SBO) occurs in the CSM and whose explosion is several tens 
of days before the SBO \cite{mor+14}, but it also fully
covered SN~2017hcd's optical flare as well. Within T$_{\rm win}$, neutrino 
events were expected to cluster in time, and we set two time profiles: 
a Gauss and a box. The neutrino flux $\Phi_{\nu_\mu+\bar\nu_\mu}$ was assumed 
to have a power-law energy spectrum with
index $\gamma$, $\Phi_{\nu_\mu+\bar\nu_\mu}=\Phi_0(E_{\nu_\mu}/E_0)^{-\gamma}$, 
where $E_{\nu_\mu}$ is neutrino energy, $E_0$ is fixed at 1\,TeV, and
$\Phi_0$ is the flux normalization.

\begin{figure}[t]
	\centering
	\includegraphics[width=0.9\textwidth]{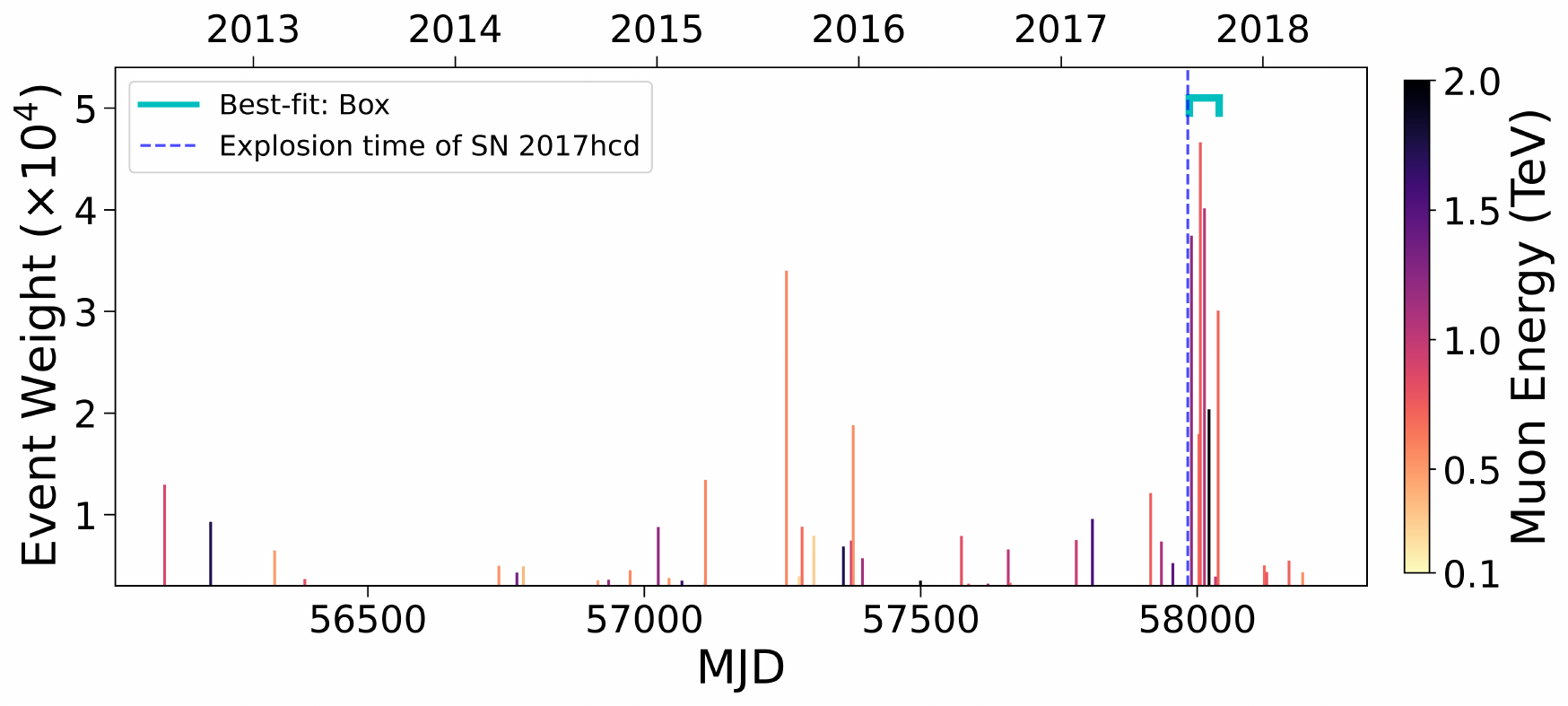}
	\caption{\textbf{Neutrino events detected at 
	SN~2017hcd's position during the IceCube season IC86-II--VII.} 
	The weights and reconstructed muon energies of the events, 
	the SN's estimated explosion time, and the neutrino flare's duration 
	from the box time profile are respectively
	indicated as the same as those in Fig.~\ref{fig:lc}.
	A clustering of high-weighted events at the SN explosion time is
	clearly visible.}
	\label{fig:weight}
\end{figure}

\begin{figure}[t]
	\centering
	\includegraphics[width=0.6\textwidth]{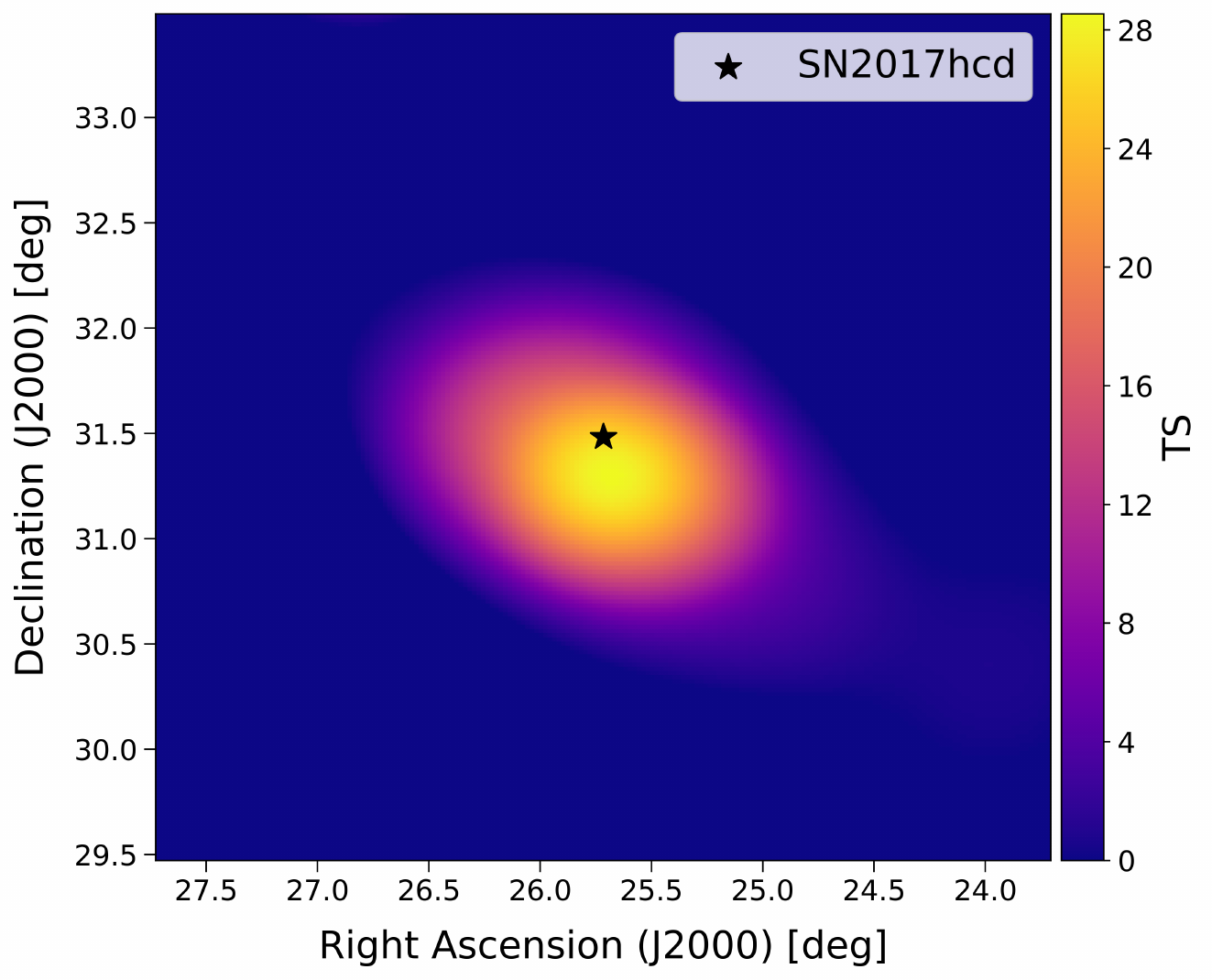}
	\caption{\textbf{Neutrino TS map of a $4^\circ\times4^\circ$ region
	around SN 2017hcd (whose position is marked by a black star)
	during the best-fit flare period.} 
	The map was calculated using the box time profile, and the spectral 
	index and time PDF were fixed at the best-fit values
	(see section `IceCube data analysis'). }
	\label{fig:tsmap}
\end{figure}

From the search analysis, we found a significant neutrino flare at
SN~2017hcd's position with both time profiles (Table~\ref{tab:neu}
and Fig.~\ref{fig:lc}). 
The box profile resulted in a higher significance value, 3.9$\sigma$, estimated 
from the obtained test statistic (TS) value of 26.3
after considering a trial factor of 2 
(ref.~\citenum{txs0506a}; see section `IceCube data analysis').
The flare, with the determined central time $T_0$ at $\sim$MJD~58013,
preceded the optical emission of the SN by $\sim$10\,days and had
a time duration $T_W$ of
approximately 34 (from the Gaussian profile) to 54 days (from
the box profile). The flare consisted of an event excess $n_s$
of 7--17 muon neutrinos in addition to the atmospheric and diffuse 
cosmic background neutrinos. 
We also calculated the weights of the long-term events at the position of 
the SN, and an event clustering at the SN's explosion time is clearly
observed (Fig.~\ref{fig:weight}).
We further obtained a TS map during the neutrino flare period 
(i.e., $T_W$ from the box time profile) by scanning 
a $4^\circ\times4^\circ$ sky region 
centered at SN~2017hcd (Fig.~\ref{fig:tsmap}), and
SN~2017hcd was only $\sim0.18^\circ$ away from the hottest location of
the neutrino emission.




The detection of a high-energy neutrino flare from an SN IIn matches the
theoretical expectations \cite{mur+11,ksw12,mur18}. A shock arising from
the collision of the SN ejecta with a CSM accelerates cosmic rays (CRs),
which collide with the CSM nucleons ($pp$ interaction), leading to 
the production of neutrinos and high-energy $\gamma$-rays. It is particularly
notable that the neutrino flare started at the estimated explosion time of 
SN~2017hcd and approximately ended after the shock breakout in the CSM
(Fig.~\ref{fig:lc}; see `Photometric data and light-curve analysis'). 
We searched for $\gamma$-ray emissions from the SN by analyzing the data
obtained with the Large Area Telescope (LAT) onboard {\it the Fermi Gamma-ray
Space Telescope}. No emissions were detected (resulting in an upper limit
on the flux in 0.1--500\,GeV, 
$\leq 8.4\times 10^{-12}$\,erg\,cm$^{-2}$\,s$^{-1}$; 
see section `{\it Fermi}-LAT data analysis'). The non-detection could 
be due to attenuation on matter and radiation in 
the CSM \cite{mur+11,mar+14,ack+15}.
However, the estimated isotropic neutrino luminosity 
at 1\,TeV (from the box profile)
was $L_{\nu,\rm iso}\simeq 3.6\times10^{45}~\rm erg~ s^{-1}$ and
the all-flavor isotropic neutrino energy at 1\,TeV was 
$E_{\nu,\rm iso}\simeq L_{\nu,\rm iso}T_W/(1+z)\approx 1.7\times10^{52}$\,erg, 
too high to be explained by the ejecta--CSM interaction scenario, as
the total radiation energy estimated from the optical light curves
was only $\sim 10^{50}$\,erg
(see section `Photometric data and light-curve analysis'). The latter
is considered to indicate the order-of-magnitude value for
the SN's kinetic energy, $\sim$10\% of which would go 
into CRs' energy \cite{mur+11,mar+14,mur18}.
The energy mismatch has been reflected in theoretical studies that only CCSNe 
(including SNe IIn) within 10\,Mpc are
shown to have potentially detectable neutrino emissions \cite{mur18,km23,vo23}.


Another widely discussed scenario for the cause of high-energy neutrino 
emissions
from CCSNe are choked jets \cite{mw01,rmw04}, which are considered
to be launched in the same manner as those of gamma-ray bursts (GRBs),
only slower than the latter. They can not break through either the envelope of
a progenitor star or a pre-existing dense CSM \cite{nak15,he+18}. 
The shock in such a jet
accelerates protons, which interact with photons ($p\gamma$ interaction)
and/or protons, producing high-energy neutrinos \cite{ab05,mi13}. 
The relativistic
beaming of this jet can significantly boost the observed neutrino 
flux \cite{ice+11,abb+12}, similar to the mechanism through which emissions 
are boosted 
in GRBs; the inferred isotropic radiation energies of GRBs are typically
as high as $\sim$10$^{52}$\,erg \cite{kwcat21}. 
Thus, the neutrino flare of SN~2017hcd might arise from a choked jet, and
the beaming effect of the jet helped enable the detection.
The flare's duration may indicate that the `unseen' jet had a $\sim$1--2 month 
lifespan. Such a long-lived jet is possible if it is powered by 
the accretion of the progenitor's mass onto a newly formed black hole 
in an SN \cite{he+18,liu+19}. 
The neutrino emission in this SN case was soft ($\gamma \sim 3.5$), comparable
to that detected from the nearby active galaxy NGC~1068 \cite{ngc1068}.
Such soft emissions have been discussed to reflect the strong cooling effects
in the emission region \cite{ab05,abb+12}. Analogous to the role of NGC~1068
(ref.~\citenum{mkm20}),
this SN case
likely shows that `hidden' jets of CCSNe are another type of high-energy
neutrino source in the Universe.



\section*{References}
\bigskip
\bigskip

\bibliographystyle{naturemag}

\clearpage

\clearpage

\clearpage

\begin{table}
    \centering
    \begin{tabular}{lccccccc}\hline
	    Profile & $T_0$ & $T_W$ & $\gamma$ & $n_s$ & $\Phi^{1\rm TeV}_{\nu_\mu+\bar\nu_\mu}/10^{-10}$ & TS & Significance \\
	    & (MJD) & (MJD) & & & (TeV$^{-1}\,$cm$^{-2}$\,s$^{-1}$) & & \\
        \hline
	    Gauss &  58014$^{+9}_{-7}$ & 34$^{+30}_{-6}$ & $3.5^{+0.5}_{-0.6}$ &$11^{+5}_{-4}$ & 3.5$^{+1.5}_{-1.2}$ & 20.2 & 3.9$\sigma$ \\
	    Box & 58012.9 & 54.3 & $3.5^{+0.5}_{-0.6}$ & $13^{+4}_{-4}$ & 2.5$^{+0.9}_{-0.8}$ & 26.3 & 4.0$\sigma$ \\
        \hline
    \end{tabular}
	\caption{{\bf Neutrino analysis results for two time profiles.}
	Columns $T_0$ and $T_W$ are the central time and
	duration of the flare, respectively; for the Gaussian time profile, 
	$T_W$ is given as 2$\sigma_T$ (section `IceCube data analysis'). 
	Columns $n_s$ and $\Phi^{1\rm TeV}_{\nu_\mu+\bar\nu_\mu}$ are 
	the net neutrino number
	and the flux normalization at 1\,TeV over $T_W$, respectively.
	Uncertainties are at a 68\% confidence level, but the upper bound values
on $\gamma$ are limited by the setting of $\gamma \leq 4.0$ in our neutrino
data analysis.
	\label{tab:neu}}
\end{table}

\clearpage

\section*{Methods}

\renewcommand{\figurename}{Extended Data Fig.}
\renewcommand{\tablename}{Extended Data Table}
\setcounter{figure}{0}
\setcounter{table}{0}
\setcounter{footnote}{0}

\section{Photometric data and light-curve analysis}
\subsection{Photometric data}

\begin{figure}[t]
	\centering
	\includegraphics[width=0.9\textwidth]{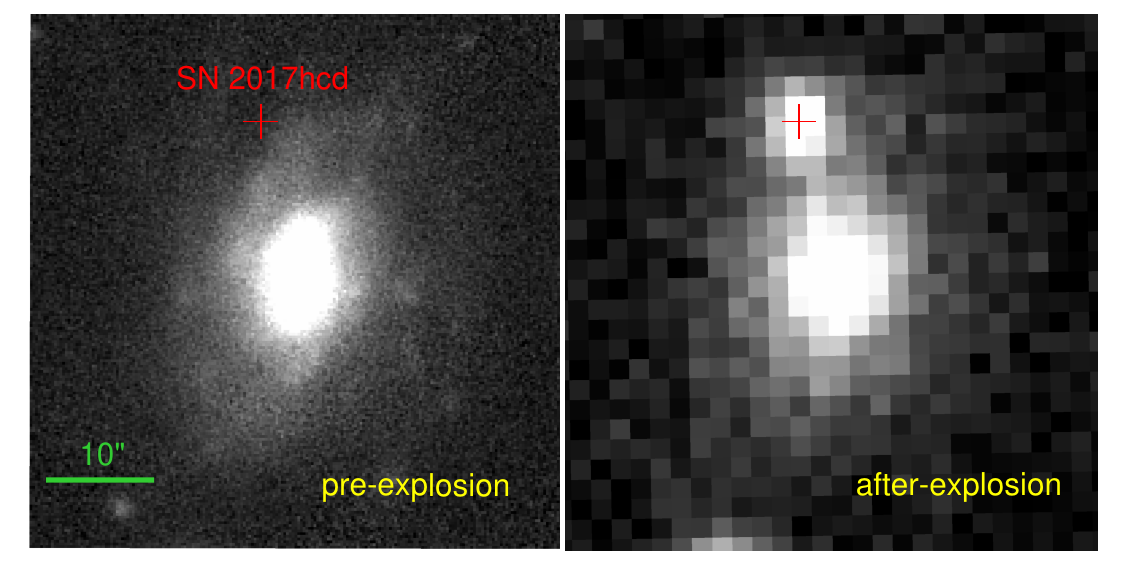}
	\caption{ \textbf{Optical images centered at the host galaxy of 
	SN~2017hcd, MCG+05-05-002.} {\it Left~panel:} Pan-STARRs $i$-band 
	image \cite{pan16} taken before 
the explosion of SN~2017hcd. {\it Right~panel:} ATLAS $o$-band image taken 
at $\sim$54 days after the discovery time. In both panels, the red cross 
	indicates the location of SN~2017hcd.}
	\label{fig:image}
\end{figure}

The discovery of SN~2017hcd was made by ATLAS on October 1 2017 at 
its $o$ band \cite{Tonrys_tns},
which was reported in the Transient Name Server 
(TNS)\footnote{\url{https://www.wis-tns.org/object/2017hcd}}. It was classified
as an SN IIn with redshift $z = 0.035$ (ref.~\citenum{Xhakaj+17}).  Its position
was at the edge of a nearby galaxy MCG+05-05-002, which has the same redshift 
value \cite{Haynes+11} and therefore is likely the host galaxy
(Extended Data Fig.~\ref{fig:image}).
Its distance to the center of the galaxy is $\sim15^{\prime\prime}$, which
corresponds to $\approx11.6$ kpc at a source distance of 159.2\,Mpc.

The ATLAS $o$-band data provide a good coverage of the entire explosion 
(see Fig.~\ref{fig:lc}). Additionally, there were a few measurements
at ATLAS $c$ band, Gaia $G$ band 
(from the Gaia Photometric Science 
Alerts \footnote{\url{https://gsaweb.ast.cam.ac.uk/alerts/alert/Gaia17dgr/}}),
Pan-STARRS $i$ band, and Sloan $g$ band. 
The ATLAS data were from
the ATLAS forced photometry server \cite{atlas21},
and were cleaned and binned for each night by using 
{\tt ATClean} \cite{atclean23}; we excluded any
data points with flux errors $>$ 0.1\,mJy. 
The non-ATLAS measurements were given on TNS. 
For the Gaia $G$-band data, the median standard deviations of different Gaia-$G$
magnitudes given in ref.~\citenum{gaiag}
were taken as the uncertainties. 
For the single Sloan $g$-band data point, we assigned 0.1 mag as its error
because no uncertainty was reported. 

\subsection{ATLAS $o$-band light curve fitting}

When an SN explodes in a dense CSM, which is sufficiently opaque to radiation, 
the shock breakout (SBO) will occur in the CSM rather than the stellar 
surface \cite{ci11}. For a CSM with a wind extent of radius
$R_w$, larger than the radiation diffusion radius $R_d$, the SBO time
$t_{\rm sbo}$ occurs at $\sim t_d=R_d/v_{\rm sh}$, where $v_{\rm sh}$ 
is the shock velocity,
and the light-curve rising time $t_{\rm rise}$ from the SBO to 
the peak is also 
$\sim t_d$ (ref.~\citenum{ci11}). The occasional flattening seen in the
declining phase of the light curve of SN~2017hcd
suggests that this SN case had $R_w>R_d$ (ref.~\citenum{ci11}).
In order to estimate the light-curve peak time, $t_{peak}$, and $t_{\rm rise}$,
we used a parabolic function \cite{ofe+14fun},
\begin{equation}
 F(t)=F_{\rm peak}[1-(\frac{t-t_{\rm peak}}{t_{\rm rise}})^2],
	\label{eq:pb}
\end{equation}
where the flux density $F(t)$ reaches the peak $F_{\rm peak}$ at time 
$t_{\rm peak}$, 
and $t_{\rm rise}$ is the duration from zero to $F_{\rm peak}$. 
To fit the rise of the ATLAS $o$-band light curve with this function, we 
applied the Markov Chain Monte Carlo (MCMC) method, implemented by Python 
package {\tt emcee} \cite{emcee}.
From the fitting, we obtained 
$t_{\rm rise}=40^{+2}_{-1}$\,day with $t_{\rm peak}$ at MJD~$58063^{+2}_{-1}$
($F_{\rm peak}=0.72 \pm 0.01$ mJy).
As the fit gave $t_{\rm sbo} \sim$ MJD~58023 when the flux was zero
and $t_d \sim t_{\rm rise} \simeq 40$\,day, 
the SN explosion time $t_{\rm exp}$
was estimated to be $\sim$40\,day before the SBO, at $\sim$MJD~57983. 
The estimated $t_{\rm exp}$ is remarkably close to the onset time of 
the neutrino flare set by the box time profile, which is at $\sim$MJD~57986
(Extended Data Fig~\ref{fig:fit}).

\subsection{Radiation energy estimate}

\begin{figure}[t]
	\centering
	\includegraphics[width=0.48\textwidth]{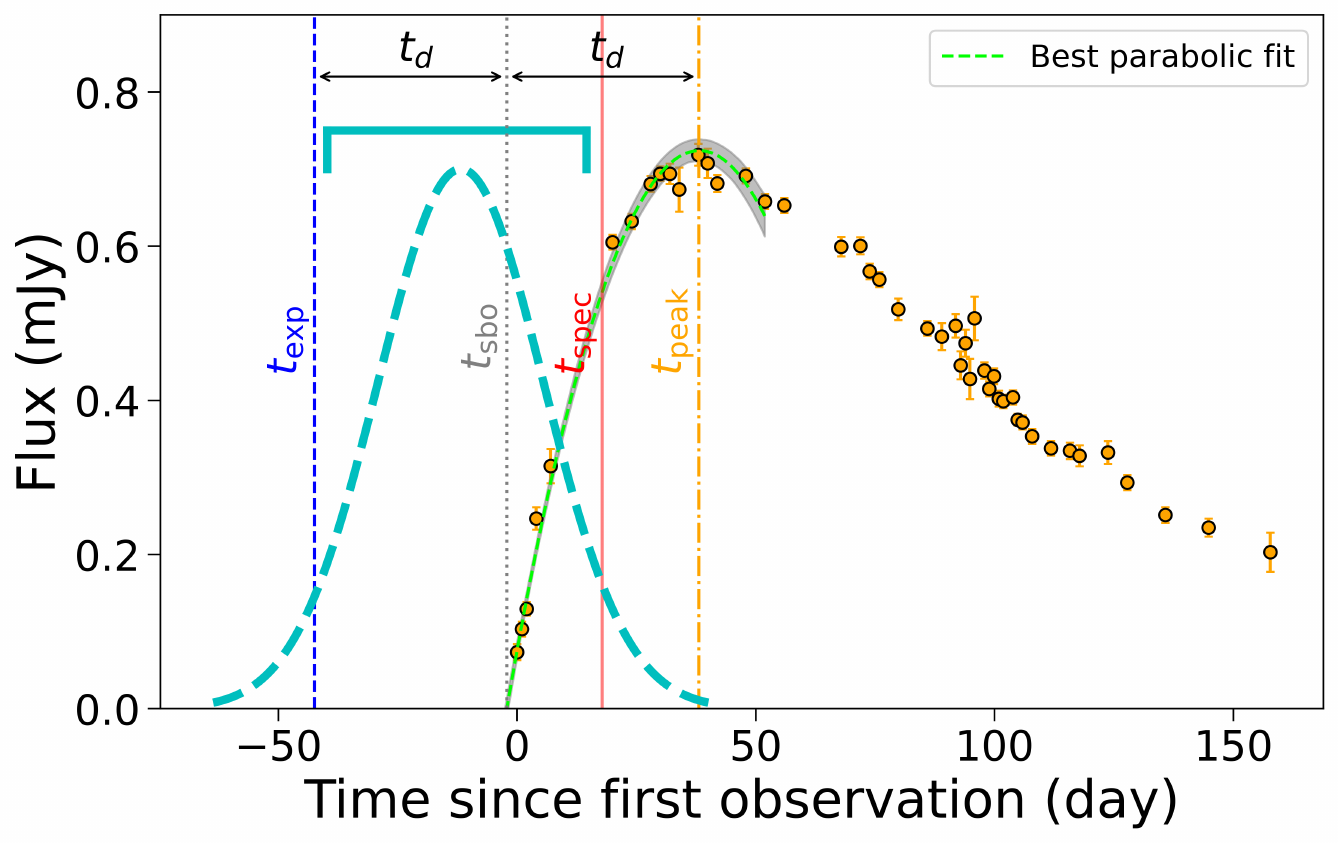}
	\includegraphics[width=0.51\textwidth]{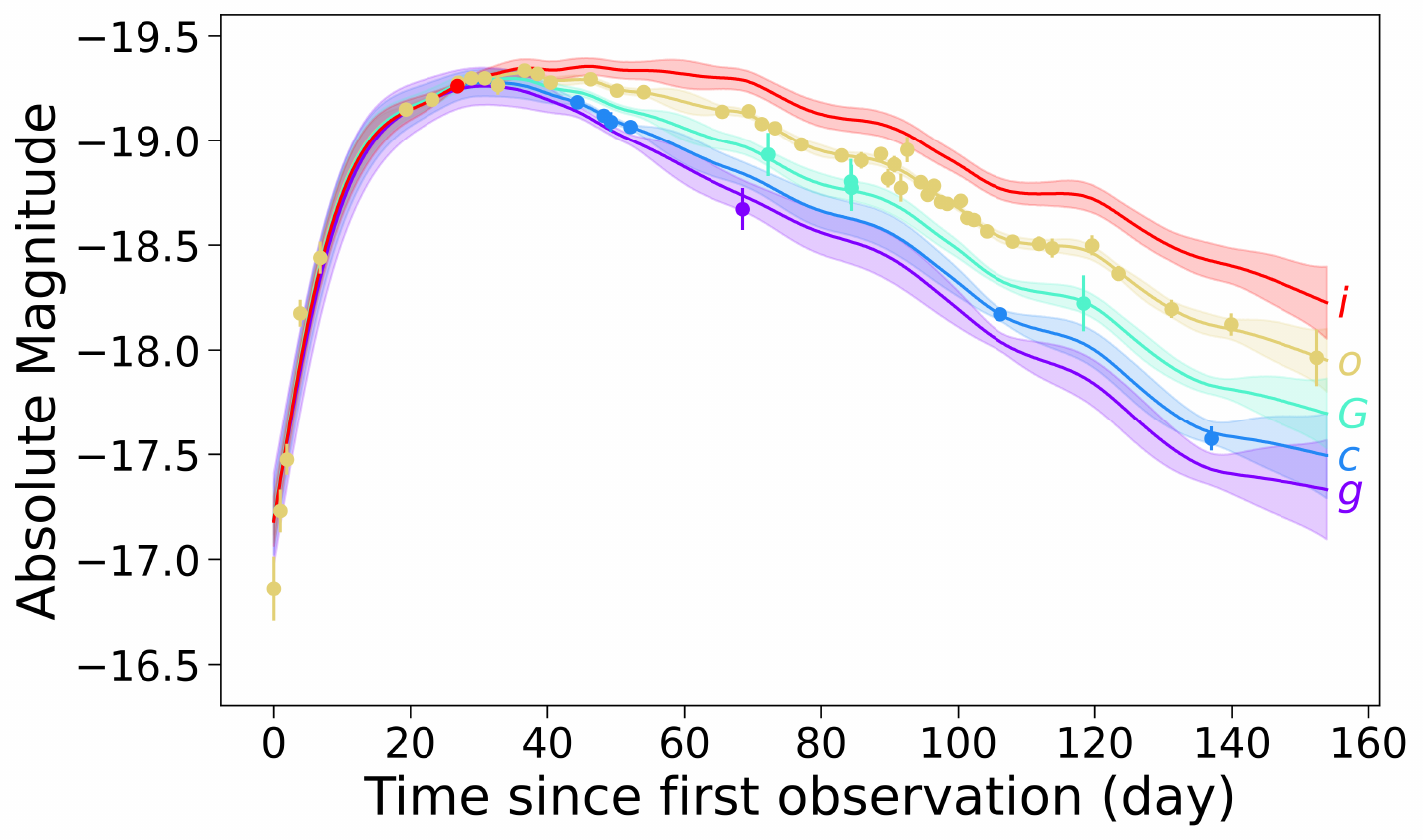}
	\caption{\textbf{\textit{Left:} parabolic fit (green dashed curve)
	to the ATLAS $o$-band 
	light curve and the estimated shock-breakout time
	(marked by the gray dotted line) in the CSM.} 
	The orange dash-dotted line indicates the flux peak time of
	the light curve, and the blue dashed line indicates the estimated 
	explosion time of SN~2017hcd. The observation time of the spectrum
	(shown in Extended Data Fig.~\ref{fig:spec}) is marked by the 
	red line. The determined two neutrino-flare time profiles 
(Fig.~\ref{fig:lc}) are also shown
	for comparison. \textbf{\textit{Right:} {\tt extrabol} fits to 
	the multi-band data points of SN~2017hcd.} The bands of the data points
	are marked at the right side of the fits.  }
	\label{fig:fit}
\end{figure}

Multi-band ultraviolet, optical, and infrared light curves of SNe IIn are 
often used to construct their bolometric light curves, which provide estimates
for their total radiation energy. However, we did not find other data except 
those reported on TNS mentioned above. In addition, the multi-band light 
curves can
be used to construct spectral energy distributions at different epochs, which
can be fitted with a blackbody (BB) function 
(e.g., ref.~\citenum{tad+13}). In order to estimate the radiation energy from
SN~2017hcd, we used a python-based package {\tt extrabol} \cite{ebol24} to 
estimate its bolometric light curve. This package uses Gaussian Process 
regression for light curve estimation
and then fits the light curves with BB functions to obtain the bolometric 
light curve. By performing
the fitting to the multi-band data (Extended Data Fig.~\ref{fig:fit}), 
in which we adopted the Galactic 
reddening $E(B-V)=0.0524$ \cite{sf11}, we obtained 
$\sim 1\times 10^{50}~\rm erg$ for the total radiation
energy from the trapezoidal integral of the estimated bolometric light curve. 
In the fitting, the temperature and radius of the BB functions were in ranges 
of 5000--9000\,K and (1--3)$\times 10^{15}$\,cm, respectively. The ranges are 
consistent with those reported for other SNe IIn (e.g., ref.~\citenum{tad+13}).

\section{Optical spectrum analysis}

\begin{figure}[t]
	\centering
	\includegraphics[width=0.7\textwidth]{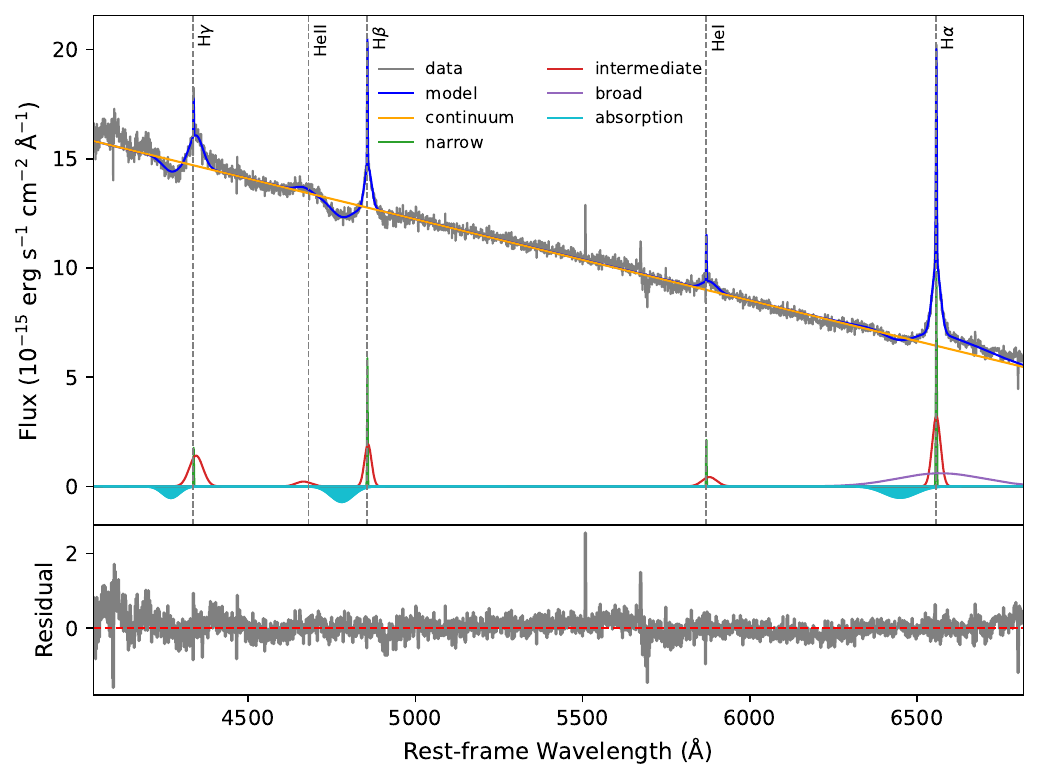}
	\includegraphics[width=0.7\textwidth]{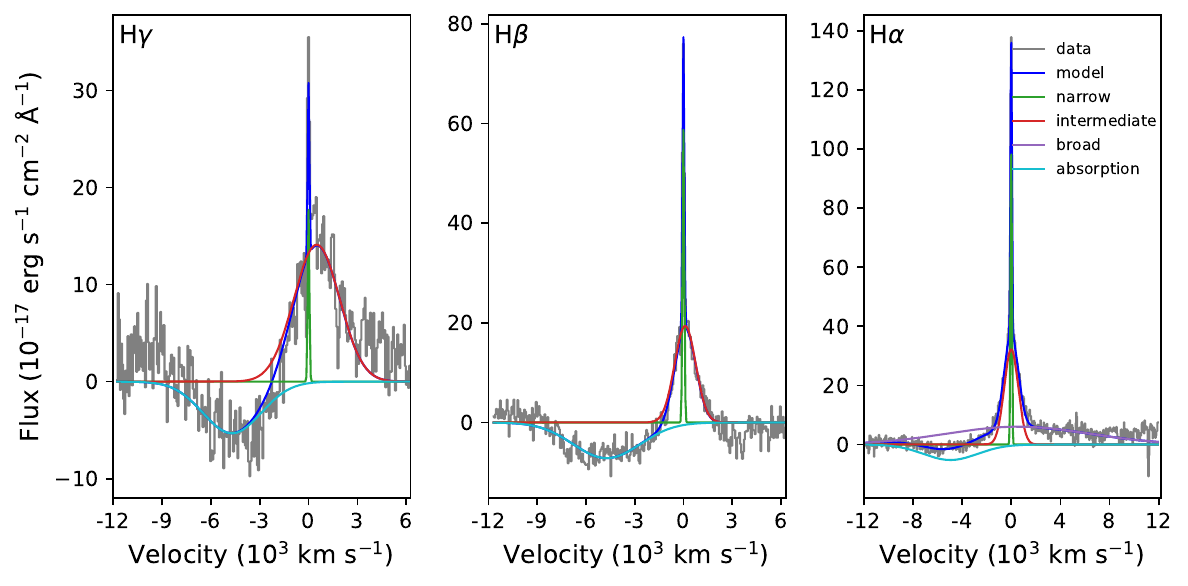}
	\caption{
		\textbf{Fitting analysis for the spectrum of SN~2017hcd.} 
		{\it Upper panels:} 
	the top shows the spectrum and the fit consisting of a linear line 
	plus multi-component Gaussian functions, and
	the bottom displays the residuals (data minus model) of the fitting. 
	{\it Lower panels:} detailed multi-component fits to the Balmer 
	H lines (H$\gamma$, H$\beta$, and H$\alpha$), whose continuum 
	component is subtracted. }
	\label{fig:spec}
\end{figure}

SN~2017hcd was classified as an SN~IIn using a spectrum taken on October 17
2017 \cite{Xhakaj+17}. We obtained this single-epoch spectrum from TNS. The 
spectrum is characterized by a blue continuum, hydrogen-rich features, and 
prominent narrow Balmer emission lines. That last characteristic is what
leads to the SN~IIn classification \cite{sch90,fil97} 
(see Extended Data Fig.~\ref{fig:spec}). We verified this classification by
using the Supernova Identification code (\texttt{SNID}) \cite{bt07}, which 
returned SN~IIn as the best match. 

We further analyzed the line features by fitting the continuum with a linear 
line and modeling the features with multi-component Gaussian 
profiles; the latter consisted of narrow, intermediate, and broad components. 
The best-fit parameters and their uncertainties were obtained using 
\texttt{emcee}. The overall fit to the spectrum 
and the detailed fits to the Balmer lines (H$\alpha$, H$\beta$, and H$\gamma$)
are presented in Extended Data Fig.~\ref{fig:spec}, and the velocity and
Full Width at Half Maximum (FWHM) measurements of the lines
are summarized in Extended Data Table~\ref{tab:hlines}. The observed spectral 
features, namely a P-Cygni profile with multi-component emissions, are 
similar to those seen in other SNe~IIn 
(e.g., refs.~\citenum{tjg+22,hmb+24,bcm+25}).
Among the components, the blue-shifted absorption feature is a direct tracer 
of the outflowing SN-CSM shell. For SN~2017hcd, the velocities from 
the absorptions at 
the Balmer lines were all consistently found within the range of 
4700--5000~km~s$^{-1}$, which was likely the expansion velocity of 
the SN-CSM shell.

\setcounter{table}{0}
\begin{table}
	\centering
	\begin{tabular}{l cc cc}
		\hline
		Name & \multicolumn{2}{c}{Absorption} & \multicolumn{2}{c}{Narrow} \\
		\phantom{H$\gamma$} & Velocity & FWHM & Velocity & FWHM \\
		\hline
		H$\alpha$ & $-4956^{+108}_{-108}$ & $5259^{+330}_{-314}$ & $-8.60^{+0.70}_{-0.60}$ & $133.4^{+1.9}_{-1.8}$ \\
		H$\beta$ & $-4729^{+74}_{-73}$ & $4933^{+157}_{-149}$ & $-2.0^{+1.3}_{-1.3}$ & $127.5^{+3.0}_{-2.9}$ \\
		H$\gamma$ & $-4700^{+20}_{-20}$ & $4265^{+172}_{-213}$ & $1^{+22}_{-20}$ & $117.3^{+6.1}_{-5.7}$ \\
		\hline
		Name & \multicolumn{2}{c}{Intermediate} & \multicolumn{2}{c}{Broad} \\
		\hline
		H$\alpha$ & $-6.7^{+6.6}_{-6.6}$ & $1342^{+24}_{-24}$ & $540^{+370}_{-353}$ & $13645^{+589}_{-557}$ \\
		H$\beta$ & $106^{+14}_{-14}$ & $1532^{+37}_{-37}$ & -- & -- \\
		H$\gamma$ & $500^{+20}_{-20}$ & $3300^{+170}_{-164}$ & -- & -- \\
		\hline
	\end{tabular}
	\caption{\textbf{Fitting Results for Balmer Lines.} Parameters were 
	derived from the multi-component Gaussian fits 
	(Extended Data Fig.~\ref{fig:spec}). The reported 
	velocities and FWHMs are in units of km~s$^{-1}$, and
	the uncertainties are 1$\sigma$ ranges.
	\label{tab:hlines}}
\end{table}
\clearpage

\clearpage

\section{IceCube data analysis}
\subsection{IceCube data and model setup}

IceCube track-like neutrino events, originating in charged-current interactions
of muon (antimuon) neutrinos with nucleons, are better suited to search for 
neutrino emissions from point-like sources because of their good angular 
resolution ($\lesssim 1^\circ$) \cite{2020PhRvL.124e1103A}. IceCube has 
released the all-sky track-like neutrino events from April 2008 to July 
2018 \cite{icecube_dataset}, which consist of five different data acquisition 
periods (seasons): IC40, IC59, IC79, IC86-I and IC86-II--VII. This 10-year 
dataset has been widely used for searching for neutrino emissions in both 
time-integrated \cite{2020PhRvL.124e1103A,2021PhRvD.103l3018Z,2022ApJ...938...38A,2022MNRAS.514..852H,2024PhRvL.132j1002N,2025ApJ...980..164L,2025ApJ...990..142J} 
and time-dependent \cite{2021ApJ...920L..45A,2023A&A...672A.102L,Abbasi+24,Li+24} analysis. 
The time-dependent analysis should only be performed on the data of one season 
because different seasons have different detector configurations.
We selected the data of season IC86-II--VII, which were obtained between 
April 2012 to July 2018, fully encompassing the optical data period of 
SN 2017hcd. 
We employed a time-dependent maximum likelihood method \cite{Braun+10}, 
implemented in {\tt SkyLLH} released by IceCube \cite{skyllh1,skyllh2,skyllh3},
to perform the neutrino data analysis.

We assumed a power-law energy spectrum for the neutrino emission,
$\Phi_{\nu_\mu+\bar\nu_\mu}=\Phi_0(E_{\nu_\mu}/E_0)^{-\gamma}$.
The flux normalization $\Phi_0$ is determined by two free parameters,
the number of signal events $n_s$ and power-law index $\gamma$, as well as
the effective area of the IceCube detector at the declination of the source
and the duration of a flare. We set up
two time profiles to search for a possible neutrino clustering:
a Gaussian and a box profile. The Gaussian profile $\mathcal{T}_{\rm S}$ is
defined as
 \begin{equation}
 \mathcal{T}_{\rm S}(t_i | T_0, \sigma_T)=\frac{1}{\sqrt{2\pi}\sigma_T}e^{-\frac
{(t_i-T_0)^2}{2\sigma_T^2}},
  	\label{eq:gauss}
  \end{equation}
where $t_i$ is the arrival time of neutrino event $i$, and $T_0$ and 
$\sigma_T$ are the central time and the standard deviation of
this Gaussian function, respectively. The time width $T_W$
for such a neutrino clustering is given as $T_W = 2\sigma_T$.
The box profile is given by a uniform distribution
  \begin{equation}
	\mathcal{T}_{\rm S}(t_i | T_0, T_W)=\frac{1}{T_W}~~(T_0-T_W/2 \leq t_i \leq T_0+T_W/2),
	\label{eq:box}
\end{equation}
where $T_0$ and $T_W$ are the central time and the time width of 
this distribution.
In our search,
$T_0$ and $\sigma_T$ (or $T_W$) were also free parameters.

\subsection{Time-dependent analysis}

The {\tt SkyLLH} document \footnote{\url{https://github.com/icecube/skyllh/blob/master/doc/user_manual.pdf}}
provides a full description about the analysis method. Here, we briefly 
introduce the process in our analysis.  

The time-dependent likelihood function is defined as:
\begin{equation}
	\mathcal{L} \left(n_s, \gamma, T_0, T_W \right) = \prod_{i} \left[ \frac{n_s}{N} S_i  \left( \vec{x}_i, \sigma_i, E_i, t_i | \vec{x}_S, \gamma, T_0, T_W\right) + \left( 1 - \frac{n_s}{N} \right) B_i  \left(\delta_i, E_i, t_i \right) \right],
	\label{eq:likelihood}
\end{equation}
where $N$ is the total number of events in the data sample, and $S_i$ and 
$B_i$ are the signal and background probability density functions (PDFs) 
for $i$th neutrino event respectively. 

The signal PDF is divided into the spatial, energy, and time components
\begin{equation}
 S_i  \left( \vec{x}_i, \sigma_i, E_i, t_i | \vec{x}_S, \gamma, T_0, T_W\right)=\mathcal{S}_{\rm S}(\vec{x}_i, \sigma_i | \vec{x}_S) \cdot \mathcal{E}_{\rm S}(E_i | \delta_i, \gamma) \cdot \mathcal{T}_{\rm S}(t_i | T_0, T_W),
	\label{eq:S_pdf}
\end{equation}
where $\mathcal{S}_{\rm S}$ is the point-spread-function (PSF) of the detector 
for a given source at position $\vec{x}_S$, which is described by a 
two-dimensional Gaussian function 
$\exp(-\frac{|\vec{x}_i - \vec{x}_S|^2}{2\sigma_i^2})$, 
with $\sigma_i$ being 
the angular reconstruction uncertainty of event $i$ at position $\vec{x}_i$. 
The signal energy PDF $ \mathcal{E}_{\rm S}$ is the probability of detecting 
a neutrino event with reconstructed muon energy $E_i$ at declination $\delta_i$ 
for an assumed power-law model. The signal time PDF $\mathcal{T}_{\rm S}$
describes the temporal distribution of neutrino events and was set to be
one of the two time profiles defined above.

The background PDF is similarly separated as
  \begin{equation}
	B_i  \left( \delta_i, E_i, t_i \right)=\mathcal{S}_{\rm B}(\delta_i) \cdot \mathcal{E}_{\rm B}(E_i | \delta_i) \cdot \mathcal{T}_{\rm B}(t_i).
	\label{eq:B_pdf}
\end{equation}
The background spatial term $\mathcal{S}_{\rm B}$ depends only on event 
declination $\delta_i$, as the background distribution is uniform in right 
ascension for IceCube. The background energy PDF $\mathcal{E}_{\rm B}$ 
describes the probability for a background event and is derived directly 
from the experimental data. The background time PDF $\mathcal{T}_{\rm B}$ is 
considered to be uniform, with 
$\mathcal{T}_{\rm B}(t_i) = \frac{1}{T_{\rm live}}$, 
where $T_{\rm live}$ is the total livetime of the data. 

For an event, its weight is defined as the signal-over-background ratio,
\begin{equation}
	S_i/B_i = \frac{\mathcal{S}_{\rm S}(\vec{x}_i, \sigma_i | \vec{x}_S) \cdot \mathcal{E}_{\rm S}(E_i | \delta_i, \gamma)}{\mathcal{S}_{\rm B}(\delta_i) \cdot \mathcal{E}_{\rm B}(E_i | \delta_i)},
	\label{eq:weight}
\end{equation}
which is time-independent and is only a function of $\gamma$ for a given 
source position. For example, the weights of the neutrino events shown in
Figs.~\ref{fig:lc} \& \ref{fig:weight} were calculated by using $\gamma = 3.5$,
the best-fit value obtained in our analysis.

In our analysis, the best-fit values were found by maximizing 
the log-likelihood ratio, which is defined as the test statistic (TS) 
\begin{equation}
	\text{TS} = 2~[\log\frac{\mathcal{L} \left(n_s, \gamma, T_0, T_W \right)}{\mathcal{L}(n_s=0)}-\log({\rm pf})],
	\label{eq:ts}
\end{equation}
where $\mathcal{L}(n_s=0)$ represents the null (background-only) hypothesis. 
A penalty factor (pf), 
$\log({\rm pf}) = \log (\frac{T_{\rm win}}{\sqrt{2\pi}\sigma_T})$ 
for a Gaussian profile ($\log (\frac{T_{\rm win}}{T_W})$ for a box profile), 
is added to correct 
the look-elsewhere effect \cite{Braun+10,Abbasi+24}. 
The $a~priori$ search window $T_{\rm win}$ was set to be 300 days,
from MJD 57950 to MJD 58250. 

\subsection{Expectation Maximization method for finding flares}

Expectation Maximization (EM) is an unsupervised learning algorithm \cite{EM} 
that has been applied in IceCube time-dependent analysis to identify neutrino 
flares \cite{Karl:2021xF,Karl:2022qgz,Abbasi+24,2024JCAP...07..057K}. This 
method can largely reduce computational cost while achieving results that are 
consistent with those obtained through conventional brute-force 
methods \cite{Abbasi+24,2024JCAP...07..057K}. It is particularly efficient 
for cases of estimating the significance of a signal when a large number of 
background trials are conducted. {\tt SkyLLH} currently only provides the EM 
fitting for a Gaussian time profile. We extended the EM fitting to 
include a box time profile. The procedure follows that given in 
refs.~\citen{Abbasi+24,2024JCAP...07..057K}.

The EM algorithm iterates between two steps: the expectation (E) step and 
the maximization (M) step. The E step calculates the membership 
probability $P_{i, \rm flare}$ for event $i$ to belong to the signal 
distribution. For a Gaussian time profile, this probability is given as
\begin{equation}
	P_{i, \text{flare}} = \frac{n_\text{flare}\frac{S_i}{B_i}\mathcal{T}(t_i|T_0, \sigma_T)}{n_\text{flare}\frac{S_i}{B_i}\mathcal{T}(t_i|T_0, \sigma_T)+\frac{N_{\text{data}}-n_\text{flare}}{T_\text{data}}},
	\label{eq:prob}
\end{equation}
where $n_\text{flare}$ denotes the expected number of signal events used 
only within the EM framework, $\frac{S_i}{B_i}$ is the time-independent 
event weight for a given spectral index $\gamma_{\rm EM}$ calculated 
from Eq.~\ref{eq:weight}, and $N_{\rm data}$ 
is the number of events in an \textit{a priori} window, whose
time length is $T_{\rm data}$.
For the box time profile, the time PDF is replaced 
by $\mathcal{T}(t_i | T_0, T_W)$, i.e., Eq.~\ref{eq:box}. In the M step, 
$T_0$ and $\sigma_T$ (or $T_W$) are updated using the membership probabilities 
calculated in the E step. For both time profiles, $T_0$ is given by
 \begin{equation}
 	T_0  = \frac{\sum\nolimits_{i}P_{i, \text{flare}}t_i}{\sum\nolimits_{i}P_{i, \text{flare}}}.
 	\label{eq:mean_t}
 \end{equation}
For the Gaussian profile, $\sigma_T$ can be directly calculated as
 \begin{equation}
	\sigma_T  = \sqrt{\frac{\sum\nolimits_{i}P_{i, \text{flare}}(t_i-T_0)^2}{\sum\nolimits_{i}P_{i, \text{flare}}}}.
	\label{eq:width_gauss}
\end{equation}
For the box profile, which is a uniform distribution, the time width can be derived as
 \begin{equation}
	T_W  = 2\sqrt{3}\sqrt{\frac{\sum\nolimits_{i}P_{i, \text{flare}}(t_i-T_0)^2}{\sum\nolimits_{i}P_{i, \text{flare}}}}.
	\label{eq:width_box}
\end{equation}
The convergence criterion is set as either the likelihood remaining unchanged 
for 20 iterations or reaching a maximum number of 500 
iterations \cite{Abbasi+24}. In our analysis, EM was performed on the data 
within the $a~priori$ window to find a neutrino flare. In addition, 
the time width $\sigma_T$ (or $T_W$) was restricted 
to $\geq$ 5 days \cite{Abbasi+24}. Upon convergence, EM returns the optimized 
$T_0$ and $\sigma_T$ (or $T_W$) for a fixed $\gamma_{\rm EM}$. 

For the box profile, the likelihood is particularly sensitive to small 
variations in time due to the discontinuity of the profile. 
If the true signal distribution is not perfectly uniform (e.g., asymmetric), 
$T_0$ and $T_W$ derived from Eq.~\ref{eq:mean_t} and Eq.~\ref{eq:width_box} 
will deviate from the actual values. Thus, we performed an additional local 
optimization of the box parameters in the vicinity of the optimized results 
given by EM. Specifically, we minimized the negative likelihood,
where the EM results serve as the initial values and the bounds are set 
as $T_0\in[T_{0, \rm EM}-T_{W, \rm EM}/2, T_{0, \rm EM}+T_{W, \rm EM}/2]$ and $T_W\in[0.5T_{W, \rm EM}, 2T_{W, \rm EM}]$.

We ran the above procedures for $\gamma_{EM}$ in the 
range [1.5, 4.0] \cite{2024JCAP...07..057K} with steps of 0.1. 
For each $\gamma_{\rm EM}$, the optimized time PDF was passed to 
Eq.~\ref{eq:ts} to maximize the likelihood ratio, yielding 
the best-fit $n_s$ and $\gamma$ values. The flare with the highest TS value 
was then identified as the best-fit flare. 
Overall, the parameters in the analysis were set in the ranges of 
$n_s\geq0, \gamma\in[1.5, 4.0], T_0\in[57950, 58250]$, and 
$T_W$ (or $\sigma_T$)$\geq5~\rm day$.
Because of the upper bound of 4.0 set on $\gamma$, the upper bound 
uncertainty for it was thus limited to 4.0 (Table~\ref{tab:neu}).

\subsection{Detection significance estimate}

\begin{figure}[t]
	\centering
	\includegraphics[width=100mm]{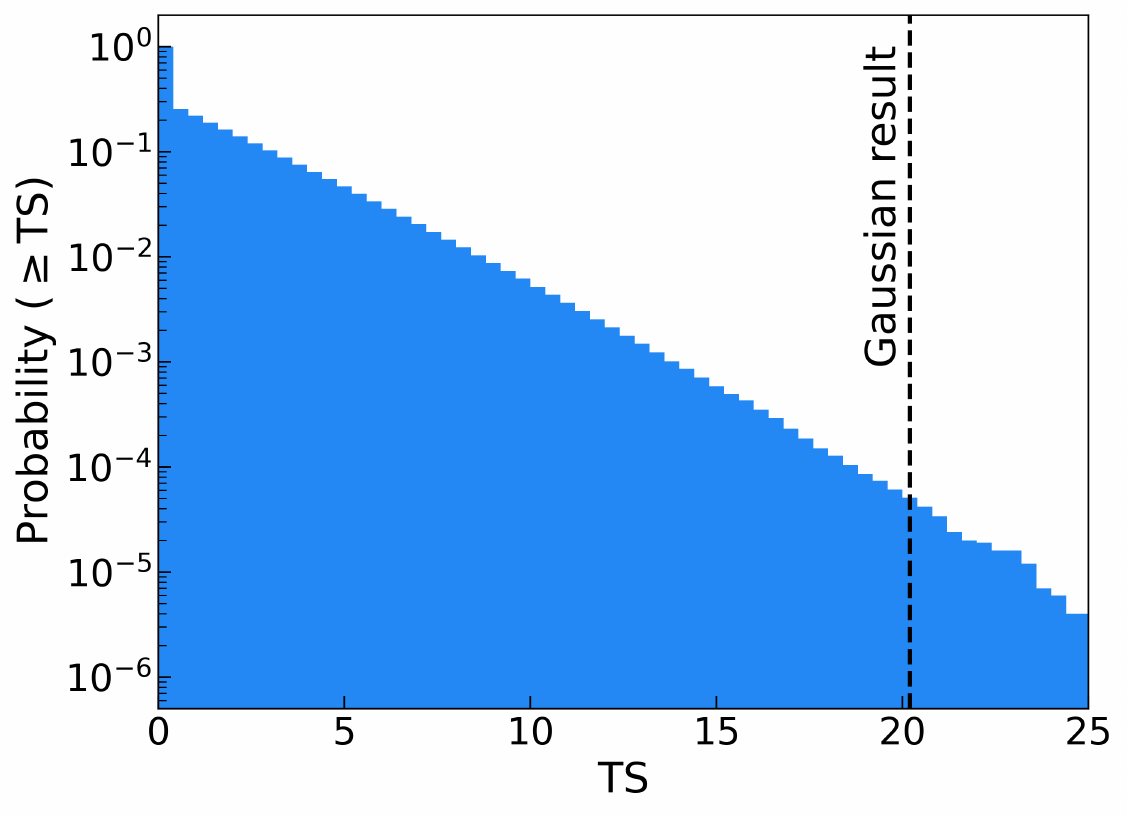}
	\includegraphics[width=100mm]{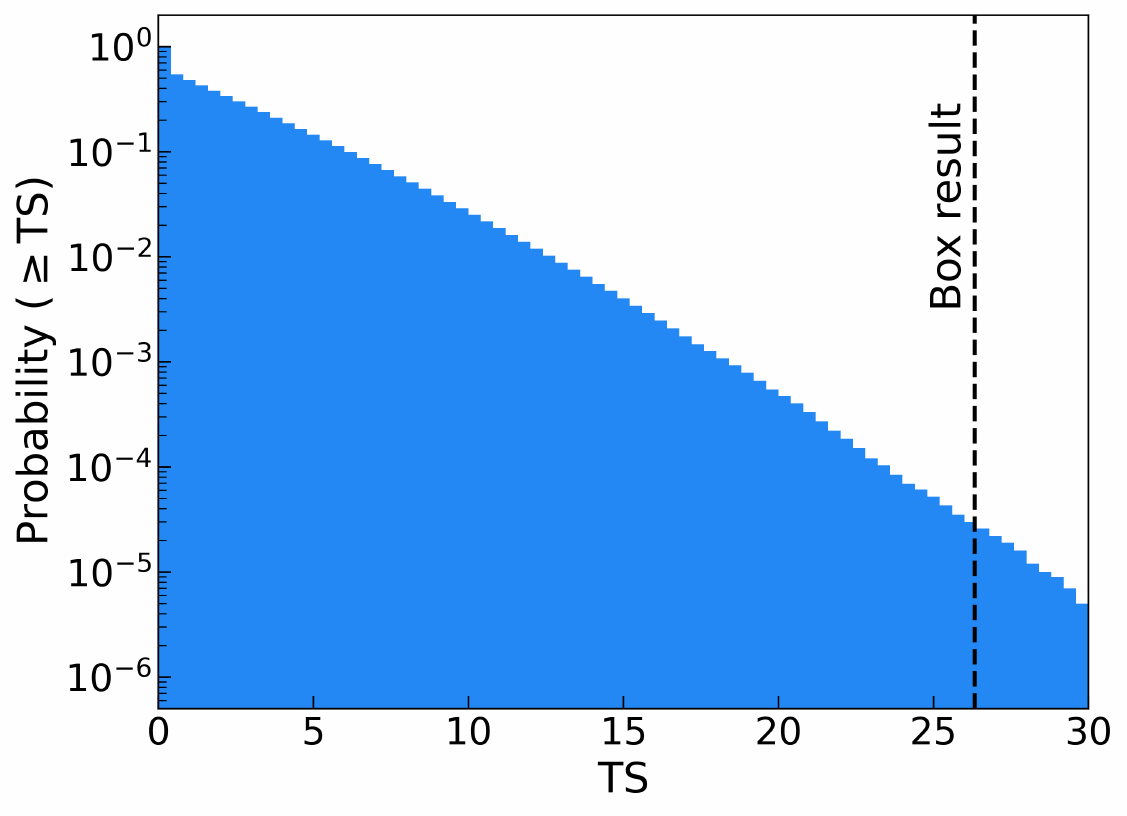}
	\caption{\textbf{Background TS distributions of the Gaussian
	(\textit{upper}) and box (\textit{lower}) time profile.} 
	The blue bar is the p-value of the background in each TS bin.
	The dashed lines mark the TS 
	values obtained at the location of SN~2017hcd with the respective 
	time profiles.
    }
	\label{fig:bkg_ts}
\end{figure}

We estimated the significance of the detection by performing background-only 
trials at the location of SN~2017hcd using randomized data. The randomized 
data were generated by shuffling the event times in the data sample and 
re-calculating the right ascension of each event. For each trial, the 
analysis following the same procedure as that applied to the real data was 
conducted.  We performed $10^6$ times background-only trials 
for each of the two time profiles.
The resulting background TS distributions are shown in Extended Data 
Fig.~\ref{fig:bkg_ts}. 
The $p$-value was then estimated as a fraction of any background trials that
had
a TS value greater than that obtained from the real-data analysis.
The largest TS values we obtained were 20.2 for the Gaussian time profile 
and 26.3 for the box time profile, and
the corresponding p-values were $4.5\times10^{-5}$ (3.9$\sigma$) for
the former and $2.7\times10^{-5}$ (4.0$\sigma$) for the latter. 
As there was no $a~priori$ time profile for the flare, we considered 
a trial factor of 2,
and the significance of the box time profile (i.e.,
the more significant one) was reduced to 3.9$\sigma$. 

\clearpage

\section{{\it Fermi}-LAT data analysis}

We searched for possible $\gamma$-ray emissions from SN~2017hcd by analyzing
the data obtained with the Large Area Telescope (LAT) onboard {\it the Fermi
Gamma-ray Space Telescope (Fermi)}.
We selected photon events in the energy range of 0.1--500 GeV 
(evclass=128 and evtype=3) from the updated {\it Fermi} Pass 8 database from
the time 
range Jul. 16 2017 00:00:00 (UTC; MJD~57950) to May 12 2018 00:00:00 
(UTC; MJD~58250), which was our search window $T_{\rm win}$ for
the neutrino flare. The region of interest (RoI) was defined 
to be $20^\circ\times20^\circ$, with the center at the position of 
SN~2017hcd.  Events 
with zenith angles $>$90$^\circ$\ were excluded to avoid contamination 
from the Earth's limb, and the expression 
DATA\_QUAL $>$ 0 \&\& LAT\_CONFIG = 1 was used to obtain good time-intervals. 
The instrumental response function P8R3\_SOURCE\_V3  and 
the package {\tt Fermitools-2.2.0} were used in our analysis.

The source model was constructed from the latest (Data Release 4, DR4)  
{\it Fermi} Gamma-ray LAT (FGL) 14-year source catalog 
(4FGL-DR4) \cite{Ballet+23}. All sources in the catalog within 25$^\circ$\ of 
the target were included, and their spectral models were adopted. In our
analysis, spectral indices and normalizations of sources within 
5$^\circ$\ of the target were set as free parameters and all other 
parameters were fixed at the catalog values. The extragalactic diffuse 
emission and the Galactic diffuse emission components, i.e., the spectral 
files iso\_P8R3\_SOURCE\_V3\_v1.txt and gll\_iem\_v07.fits, respectively, were 
included as well. The normalizations of these two background components were 
also set as free parameters.

For possible $\gamma$-ray emissions from SN~2017hcd, we assumed a point source 
with a power-law energy spectrum, 
$dN/dE_{\gamma} = N{_0}(E_\gamma/E{_b})^{-\Gamma}$, 
where $E{_b}$ was fixed at 1\,GeV. We performed the standard binned likelihood 
analysis \footnote{\url{https://fermi.gsfc.nasa.gov/ssc/data/analysis/documentation/Cicerone/Cicerone_Likelihood/}} to the whole selected data in 0.1--500 GeV,
and did not find significant emission at the target's position. 
We also performed unbinned likelihood analysis to the data
in the 54-day neutrino-emission time range (Fig.~\ref{fig:lc}), and did not
find significant emission either.
Individual $\geq$1\,GeV photons from the direction of SN~2017hcd were searched
by running tool {\tt gtsrcprob} for data within $1^\circ$ of the
SN, which is approximately the 68\% containment angle 
of LAT's PSF at 1\,GeV\footnote{\url{https//www.slac.stanford.edu/exp/glast/groups/canda/lat_Performance.htm}}.
The spectral index $\Gamma$ for an assumed 
source was fixed at 2. We did
not find any photons with $\geq$50\% probability arising from the direction
of SN~2017hcd. The 95\% photon-flux upper limit during the 54-day time range
was estimated to be $\sim 6\times 10^{-9}$\,cm$^{-2}$\,s$^{-1}$ in
the energy range of 0.1--500\,GeV, which corresponds to an energy-flux upper
limit of $8.4\times 10^{-12}$\,erg\,cm$^{-2}$\,s$^{-1}$ in the energy range
of 0.1--500\,GeV (where $\Gamma = 2$ was assumed).

\section*{References}
\bigskip
\bigskip

\bibliographystyle{naturemag}

\clearpage

\begin{addendum}
 \item [Acknowledgments] This work has made use of data from the Asteroid Terrestrial-impact Last Alert System (ATLAS) project. The Asteroid Terrestrial-impact Last Alert System (ATLAS) project is primarily funded to search for near earth asteroids through NASA grants NN12AR55G, 80NSSC18K0284, and 80NSSC18K1575; byproducts of the NEO search include images and catalogs from the survey area. This work was partially funded by Kepler/K2 grant J1944/80NSSC19K0112 and HST GO-15889, and STFC grants ST/T000198/1 and ST/S006109/1. The ATLAS science products have been made possible through the contributions of the University of Hawaii Institute for Astronomy, the Queen’s University Belfast, the Space Telescope Science Institute, the South African Astronomical Observatory, and The Millennium Institute of Astrophysics (MAS), Chile.

We acknowledge ESA Gaia, DPAC and the Photometric Science Alerts Team
(\url{http://gsaweb.ast.cam.ac.uk/alerts}).
 
This research is supported by the National Natural Science Foundation of 
	China (11633007) and the Xingdian Talent Support Project of
	the Yunnan Province (XDYC-YLXZ-2023-0016).
S.J. acknowledges the support from the Yunnan Provincial Department of 
Education Scientific Research Fund and the Yunnan University Graduate Student 
Research Program (KC-252512068).
 \item[Author contributions] S.J. made the initial discovery, conducted
	the IceCube data analysis, and wrote the initial text for the Methods
	sections. Z.W. formed the concept for the discovery
	and wrote most of the text. L.Z. analyzed the optical spectrum data.
D.Z. analyzed the Fermi-LAT data.
 \item[Competing Interests] The authors declare no competing interests.
 \item[Correspondence] Correspondence and requests for materials should be 
addressed to Zhongxiang Wang (wangzx20@ynu.edu.cn).
\end{addendum}

\end{document}